\documentclass[bibyear,letter]{aa} 

\usepackage{epsfig}
\usepackage{amsmath}
\usepackage{subfigure}
\usepackage{natbib}
\usepackage{multirow}
\usepackage{hyperref}
\usepackage{longtable}
\usepackage{graphicx}
\usepackage{epstopdf}
\usepackage{booktabs}
\usepackage{stackrel,amssymb,amsmath}
\usepackage{txfonts}
\usepackage[utf8]{inputenc}%

\newcommand{\jms}{J.~Mol.~Spectrosc.}

\newcommand{\jmst}{J.~Mol.~Struct.}

\newcommand{\kms}{km s$^{-1}$}

\begin{document}

\title{More sulphur in TMC-1: Discovery of the NC$_3$S and HC$_3$S radicals with the 
QUIJOTE line survey\thanks{Based
on observations with the Yebes 40m telescope (projects 19A003, 20A014, 20D023, 21A011, 21D005,
and 23A024). The 40 m radio telescope at Yebes Observatory is operated by the Spanish Geographic 
Institute (IGN; Ministerio de Transportes y Movilidad Sostenible).
 }}

\author{
J.~Cernicharo\inst{1},
C.~Cabezas\inst{1},
M.~Ag\'undez\inst{1},
R.~Fuentetaja\inst{1},
B.~Tercero\inst{2,3},
N.~Marcelino\inst{2,3}, and
P.~de~Vicente\inst{3}
}

\institute{Dept. de Astrof\'isica Molecular, Instituto de F\'isica Fundamental (IFF-CSIC),
C/ Serrano 121, 28006 Madrid, Spain. \newline \email jose.cernicharo@csic.es
\and Observatorio Astron\'omico Nacional (OAN, IGN), C/ Alfonso XII, 3, 28014, Madrid, Spain.
\and Observatorio de Yebes (IGN). Cerro de la Palera s/n, 19141 Yebes, Guadalajara, Spain
}

\date{Received: 26/06/2024; accepted: 05/07/2024}

\abstract{
We present the detection of the free radicals NC$_3$S and HC$_3$S towards TMC-1
with the QUIJOTE line survey. The derived column densities are (1.4$\pm$0.2)$\times$10$^{11}$ for NC$_3$S and 
(1.5$\pm$0.2)$\times$10$^{11}$ for HC$_3$S. We searched for NCCS, but only three transitions are within the 
domain of our QUIJOTE line
survey and the observed lines are marginally detected at the 3$\sigma$ level, providing an upper limit to its 
column density of $\leq$6$\times$10$^{10}$ cm$^{-2}$.
We also unsuccessfully searched for longer species of the NC$_n$S (n$\ge$4) and HC$_n$S (n$\ge$5)
families in our TMC-1 data. A chemical model based on 
a reduced set of reactions involving HC$_3$S and NC$_3$S predicts abundances that are 10-100 times 
below the observed values. These calculations indicate that the most efficient reactions of formation 
of HC$_3$S and NC$_3$S in the model are S + C$_3$H$_2$ and N + HC$_3$S, respectively, while both radicals 
are very efficiently 
destroyed through reactions with neutral atoms.}


\titlerunning{HC$_3$S and NC$_3$S}
\authorrunning{Cernicharo et al.}

\maketitle

\section{Introduction}
A significant number of S-bearing species have been detected in the starless core TMC-1. In particular, the species 
CS, CCS, and CCCS are among the most abundant molecules in cold dark clouds 
\citep{Saito1987,Yamamoto1987,Kaifu1987} and in the circumstellar
envelopes of carbon-rich evolved stars \citep{Cernicharo1987a}. The abundance and variety of sulphur-bearing species detected 
to date in cold clouds is hampered, to a large extent, by the significant depletion of sulphur in the cloud \citep[][and references 
therein]{Vidal2017,Fuente2019,Navarro2020,Fuente2023} and by the lack of sensitive line
surveys that allow the detection of low-abundance, or high-abundance but low-dipole moment, S-bearing molecules.

In the last four years, our knowledge of the chemical composition of TMC-1 regarding sulphur
has experienced an enormous advancement 
with the discovery of 14 new sulphur-bearing molecules thanks to the ultra-sensitive QUIJOTE\footnote{\textbf{Q}-band 
\textbf{U}ltrasensitive \textbf{I}nspection \textbf{J}ourney
to the \textbf{O}bscure \textbf{T}MC-1 \textbf{E}nvironment} line survey \citep{Cernicharo2021a}. Neutral species such as
C$_4$S, C$_5$S (previously detected only towards evolved stars), HCCS, H$_2$CCS, H$_2$CCCS, NCS, HSO, HCSCN, HCSCCH, HC$_4$S, HCNS, and 
NCCHCS have been detected with QUIJOTE \citep{Cernicharo2021b,Cernicharo2021c,Fuentetaja2022,Marcelino2023,Cernicharo2024a,Cabezas2024}. 
In addition, cations such as HCCS$^+$ \citep{Cabezas2022} and HC$_3$S$^+$ \citep{Cernicharo2021d} have also been found in this cloud. 
Chemical models are being continuously improved and refined to shed light on the formation and destruction pathways of all these newly 
detected species.

The continuous sensitivity increase of the QUIJOTE line survey makes it well suited to detect additional heavier S-bearing 
species. In this context 
we present in this Letter the discovery, for the first time in 
space, of the radicals HC$_3$S and NC$_3$S. Only upper limits have been obtained for the column densities of longer members of the 
HC$_n$S and NC$_n$S families. The formation and destruction pathways of HC$_3$S and NC$_3$S are analysed 
with state-of-the-art chemical models. With the detection of HC$_3$S and NC$_3$S, QUIJOTE has provided the identification of 16 
new S-bearing molecules in space.

\section{Observations}

The observational data used in this work are from QUIJOTE \citep{Cernicharo2021a},
a spectral line survey of TMC-1 in the Q band carried out with the Yebes 40m telescope at
the position $\alpha_{J2000}=4^{\rm h} 41^{\rm  m} 41.9^{\rm s}$ and $\delta_{J2000}=
+25^\circ 41' 27.0''$, which corresponds to the cyanopolyyne peak in TMC-1.
The receiver
was built as part of the Nanocosmos project\footnote{\texttt{https://nanocosmos.iff.csic.es/}}
and consists of two cold high-electron mobility transistor amplifiers covering the
31.0-50.3 GHz band with horizontal and vertical polarizations.
Receiver temperatures
vary between 16\,K at 32 GHz and 30\,K at 50 GHz. The back ends are
$2\times8\times2.5$ GHz fast Fourier transform
spectrometers with a spectral resolution of 38 kHz, providing the whole coverage
of the Q band in both polarizations.
A detailed description of the system
is given by \citet{Tercero2021}, and details on the QUIJOTE line survey observations
have previously been provided \citep{Cernicharo2021a,Cernicharo2023a,Cernicharo2023b,Cernicharo2024a,
Cernicharo2024b}. The frequency switching method was used for all observations.
The data analysis procedure has been described by \citet{Cernicharo2022}.
The total observing time on source is 1202 hours, and the
measured sensitivity varies between 0.08 mK at 32 GHz and 0.2 mK at 49.5 GHz.

The main beam efficiency can be given across the Q band as
$\eta_{\rm eff}$=0.797 exp[$-$($\nu$(GHz)/71.1)$^2$]. The
forward telescope efficiency is 0.95, and
the beam size at half power intensity is 54.4$''$ and 36.4$''$
at 32.4 and 48.4 GHz, respectively.
The absolute intensity calibration uncertainty is 10$\%$, although the relative
calibration between lines within the QUIJOTE survey is certainly better because all
of them are observed simultaneously and have the same calibration uncertainties and systematic
effects.
The data were analysed with the GILDAS package\footnote{\texttt{http://www.iram.fr/IRAMFR/GILDAS}}.

\section{Results}

Line identification in this work was performed using the 
MADEX code \citep{Cernicharo2012b}, in addition to the CDMS
and JPL catalogues \citep{Muller2005,Pickett1998}.
The intensity scale
utilized in this study is the antenna temperature ($T_A^*$). Consequently, the telescope parameters and source
properties were
used when modelling the emission of the different species to produce synthetic spectra on this
temperature scale. In this work we assumed
a velocity for the source relative to the
local standard of rest of 5.83 \kms\, \citep{Cernicharo2020}. The source was assumed to be circular
with a uniform brightness temperature and a radius of 40$''$ \citep{Fosse2001}.
Line parameters for all observed transitions with the Yebes 40m radio telescope
were derived by fitting a Gaussian line profile to them
using the GILDAS package. A
velocity range of $\pm$20\,\kms\, around each feature was considered for the fit after a polynomial
baseline was removed. Negative features produced in the folding of the frequency switching data were blanked
before baseline removal.
The observed line intensities were modelled using a local thermodynamical equilibrium hypothesis.
The laboratory data used in this work are discussed in the following sections. We first discuss NC$_3$S because
its detection motivated the search for NCCS and other longer NC$_n$S chains.

\begin{table*}
\centering
\caption{Estimated line parameters for the observed lines of NC$_3$S.}
\label{line_centroid}
\begin{tabular}{lccccccc}
\hline
\hline
$J_u-J_l$    & $F_u-F_l$ & $\nu_{obs}$~$^a$    & obs.-calc.   & $\int T_A^* dv$~$^b$ & $\Delta$v$^c$ & T$_A$$^*$$^d$ & Notes\\
             &     &  (MHz)              & (kHz) & (mK km\,s$^{-1}$)    &               &  (mK)     & \\
\hline
    23/2-21/2&          & 33096.164$\pm$0.020 & 2.6   & 1.79$\pm$0.11    & 3.02    & 0.61$\pm$0.09&A\\
    23/2-21/2& 25/2-23/2& 33096.047$\pm$0.020 & 13.0   & 0.53$\pm$0.05    & 0.70$^e$& 0.71$\pm$0.09&\\
    23/2-21/2& 23/2-21/2& 33096.168$\pm$0.020 & -11.0   & 0.61$\pm$0.05    & 0.70$^e$& 0.83$\pm$0.09&\\
    23/2-21/2& 21/2-19/2& 33096.303$\pm$0.020 &  20.0   & 0.50$\pm$0.05    & 0.70$^e$& 0.66$\pm$0.09&\\
    25/2-23/2&          & 35974.029$\pm$0.020 &-2.3   & 1.73$\pm$0.11    & 2.10    & 0.78$\pm$0.08&\\
    25/2-23/2& 27/2-25/2& 35973.943$\pm$0.020 & 23.0   & 0.74$\pm$0.14    & 0.90$\pm$0.22& 0.77$\pm$0.08&\\
    25/2-23/2& 25/2-23/2& 35974.048$\pm$0.020 & 25.0   & 0.46$\pm$0.18    & 0.61$\pm$0.25& 0.76$\pm$0.08&\\
    25/2-23/2& 23/2-21/2& 35974.138$\pm$0.020 & 11.0   & 0.48$\pm$0.12    & 0.62$\pm$0.09& 0.72$\pm$0.08&\\
    27/2-25/2&          & 38851.890$\pm$0.020 & 3.0   & 1.27$\pm$0.11    & 2.20         & 0.54$\pm$0.11&\\
    29/2-27/2&          & 41729.725$\pm$0.020 &-2.2   & 0.82$\pm$0.10    & 1.20         & 0.70$\pm$0.12&\\
    31/2-29/2&          & 44607.546$\pm$0.020 &-4.9   & 0.76$\pm$0.08    & 1.39         & 0.51$\pm$0.11&\\
    33/2-31/2&          & 47485.361$\pm$0.020 & 4.1   & 0.89$\pm$0.18    & 1.30         & 0.69$\pm$0.15&A\\
\hline
\end{tabular}
\tablefoot{
\tablefoottext{a}{Observed frequency adopting a
v$_{LSR}$ of 5.83 km\,s$^{-1}$ for the source \citep{Cernicharo2020}.
If $F$ is not given, then the value refers to the frequency centroid and the calculated
frequency is obtained from $B_{eff}$ and $D_{eff}$. For lines with resolved hyperfine structure,
the calculated frequencies have been obtained from a fit to the laboratory data from 
\citet{McCarthy2003}.}
\tablefoottext{b}{Integrated line intensity in mK\,km\,s$^{-1}$.}
\tablefoottext{c}{Equivalent line width (in km~s$^{-1}$).}
\tablefoottext{d}{Antenna temperature in millikelvins.}
\tablefoottext{e}{Line width has been fixed.}
\tablefoottext{A}{Partially blended with an identified feature.}
}
\end{table*}

\begin{figure}
\centering
\includegraphics[width=0.48\textwidth]{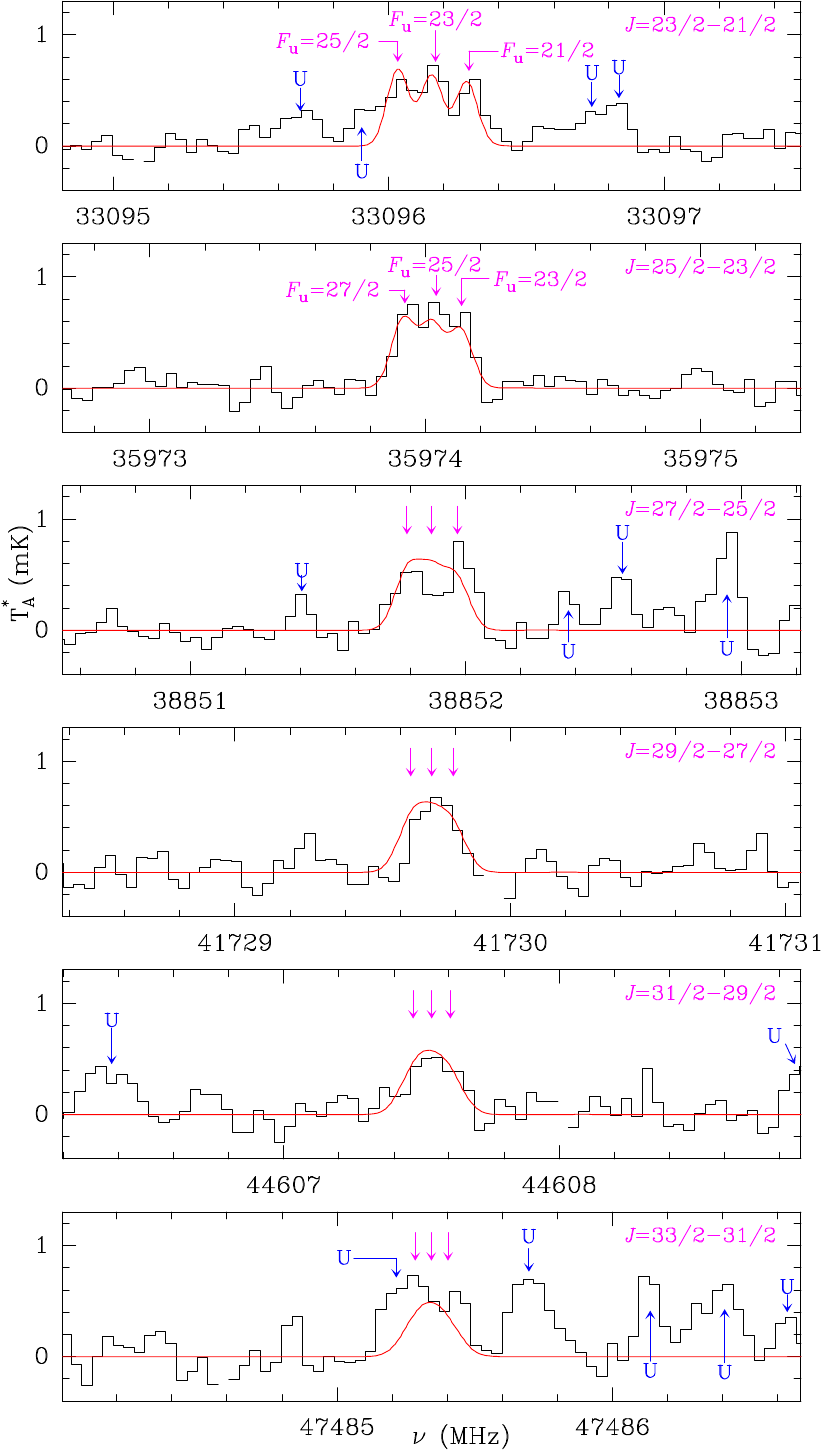}
\caption{Observed transitions of NC$_3$S. Quantum numbers are indicated at the
top right of each panel.
The abscissa corresponds to the rest frequency. The ordinate is the antenna temperature, corrected for
atmospheric and telescope losses, in  millikelvins.
Blank channels correspond to negative features produced when folding the frequency-switched data.
The purple vertical arrows indicate the position of the
three strongest hyperfine components for each transition, corresponding
to $F_u=J_u+1$, $J_u$, and $J_u-1$.
The red line shows the modelled spectra for NC$_3$S with the physical parameters given in Sect. \ref{sec:results_NCCCS}.}
\label{fig_NCCCS}
\end{figure}

\subsection{Discovery of NC$_3$S}\label{sec:results_NCCCS}

We have found a series of six harmonically related lines with half-integer quantum numbers from $J$=23/2-21/2
to $J$=33/2-31/2. The lines are shown in Fig. \ref{fig_NCCCS}. The two first
lines clearly show hyperfine structure, which suggests the presence of a nitrogen atom in the molecule. The other 
lines are also broader with respect to the standard lines from other species ($\Delta$v=0.6-0.7 km\,s$^{-1}$).
We were able to fit the frequency centroids (see Table \ref{line_centroid}) with the standard relation for the 
frequencies of a linear molecule, finding an effective rotational
constant, $B_{\rm eff}$, of 1438.9761$\pm$0.0003 MHz and an effective distortion constant, $D_{\rm eff}$, of 
47.6$\pm$0.6 Hz. The standard deviation of the fit is 4.1 kHz.
We verified that the lines cannot be produced by a species
with a rotational constant half of the
derived one (all lines with even $J$s would be missing in that case).
The half-integer quantum numbers indicate that the carrier
is a radical.

The lines do not appear in any of the line catalogues available to us. The recent discovery of HC$_5$N$^+$ 
with a rotational constant of 1336.6 MHz \citep{Cernicharo2024b} pointed us towards the possibility of an isomer
of this molecule. Accurate ab initio calculations, however, permitted us to discard all of them. 
A species
recently detected in TMC-1 is HC$_4$S \citep{Fuentetaja2022}, which has a rotational constant of 1435.3 MHz and a $^2\Pi$ inverted ground electronic state \citep{Hirahara1994}. Hence, the carrier should be a molecule 
similar to HC$_4$S but containing a nitrogen atom. Taking into account that NCS has already been detected in 
TMC-1 \citep{Cernicharo2021b}, we conclude that NC$_3$S is a solid candidate to be
the carrier of the observed lines. This species was not included in the MADEX code, nor in the CDMS or JPL catalogues. However, a quick search in the literature indicates that the molecule
has been studied in the laboratory by \citet{McCarthy2003}, together with other longer members of the NC$_n$S family.

NC$_3$S has an inverted $^2\Pi$ ground electronic state, and the transitions observed in the laboratory by \citet{McCarthy2003} cover from $J$=5/2-3/2 up to $J$=17/2-15/2, which corresponds to frequencies up to 24.46 GHz. These data were fitted using SPFIT \citep{Pickett1991}, and the molecule
was implemented in the MADEX code to produce predictions in the frequency domain of QUIJOTE. The spin-orbit constant was fixed to the value given by \citet{McCarthy2003}, $A_{SO}$=$-$9.8209 THz. With such a high value for the spin-orbit
constant, the energy levels of the $^2\Pi_{1/2}$ ladder are above 470\,K, and hence, no lines from this state
are expected in TMC-1, where the kinetic temperature is 9\,K \citep{Agundez2023}. 
The derived rotational constant is $B_0$=1439.18582$\pm$0.00008 MHz, from which an effective rotational
constant for the $^2\Pi_{3/2}$ ladder can be derived from the relation $B_{\rm eff}$=$B_0$(1+$B_0$/$A_{SO}$). 
Using the molecular constants derived by \citet{McCarthy2003}, we derive $B_{\rm eff}(^2\Pi_{3/2})$=1438.97492$\pm$0.00008 MHz, which is nearly identical 
to that derived for our series of lines in TMC-1 (a difference of 1.2 kHz). The distortion constant of NC$_3$S derived from the laboratory data
is 45.8$\pm$0.8 Hz, also in excellent
agreement with that derived for our series of lines.
Hence, we conclude that the carrier of our lines is, without any doubt, NC$_3$S. 

The lines observed in the laboratory for this species exhibit a prominent hyperfine structure introduced by the nitrogen nucleus. However, no $\Lambda$-doubling splitting was found. As an example, the three strongest components of the $J$=17/2-15/2 transition at 24.6 GHz span 0.5 MHz. The frequency predictions indicate that the $J$=23/2-21/2 and 25/2-23/2 lines observed in TMC-1 will show a considerable splitting,  as observed in our lines
(see Fig. \ref{fig_NCCCS}). All the other transitions within the
frequency range of our data are predicted to have a significant line broadening.
We merged the astronomical frequencies with the laboratory values to try to improve
the rotational constants of NC$_3$S. However, such an improvement was marginal
as the derived parameters are nearly the same as those of \citet{McCarthy2003}.

To derive the column density of NC$_3$S, we adopted a dipole moment of 3.0\,D \citep{McCarthy2003}.
We fitted the observed line profiles and intensities
assuming a source of uniform brightness temperature. 
We obtain T$_{rot}$=5.5$\pm$0.5\,K and a column density of (1.4$\pm$0.2)$\times$10$^{11}$ cm$^{-2}$. The source diameter that best fits the data is 70$''$. The observed line profiles
are nicely reproduced by the modelled synthetic spectra shown in red in Fig. \ref{fig_NCCCS}.

\subsection{Search for NCCS and longer NC$_n$S chains}\label{sec:results_NCCS}

Since NCS and NC$_3$S are detected \citep[][and this work]{Cernicharo2021b}, it was worthwhile searching for NCCS in our data. 
Similarly to NC$_3$S, the NCCS species was not included in the MADEX 
code, nor in the CDMS or JPL catalogues. However, the bent radical NCCS ($X^2A$') was studied via Fourier-transform microwave 
spectroscopy by \citet{Nakajima2003} up to
22.6 GHz, corresponding to the $4_{04}-3_{03}$ transition. Only $K_a$=0 transitions were observed
in the laboratory. The observed
frequencies were fitted using the SPFIT code \citep{Pickett1991} to produce predictions in the frequency domain
of QUIJOTE. The $A$ constant was fixed to 186.47 GHz; hence, the $K_a$=1 lines are $\sim$9\,K above the $K_a$=0 ones and will 
have a considerable frequency uncertainty. The first $K_a$=1 line in our data is the $6_{16}-5_{15}$ with an upper energy level 
at 14.4\,K. For an expected rotational temperature
of 5-6\,K, as found for another three or four atoms species, the expected intensities of the $K_a$=1 lines
will be a factor of five weaker than those of the $K_a$=0 ones. Three transitions with $K_a$=0 can be explored with our data ($N_{\rm up}$\,=\,6, 7, and 8). These lines show two fine
components with $J$=$N$+1/2 and $J$=$N$-1/2, and an additional hyperfine splitting introduced
by the nitrogen atom. The fine and hyperfine components of each rotational transition
span 200 kHz. Unfortunately, some lines are blended, and the other lines
are only detected at a marginal 3$\sigma$ level. 
The dipole moment of NCCS has been calculated to be 2.49\,D \citep{McCarthy2003}. Using this value, we derive a 3$\sigma$ upper limit to the column density of this radical
of $\leq$6$\times$10$^{10}$ cm$^{-2}$.

Laboratory data for NC$_4$S, NC$_5$S, NC$_6$S, and NC$_7$S are also available from
the study of \citet{McCarthy2003}. We implemented these molecules in MADEX and
searched for them in our data. However, none of their lines were detected. The 3$\sigma$
upper limits to their column densities are in the range (8-10)$\times$10$^{10}$ cm$^{-2}$.

\begin{figure}
\centering
\includegraphics[width=0.49\textwidth]{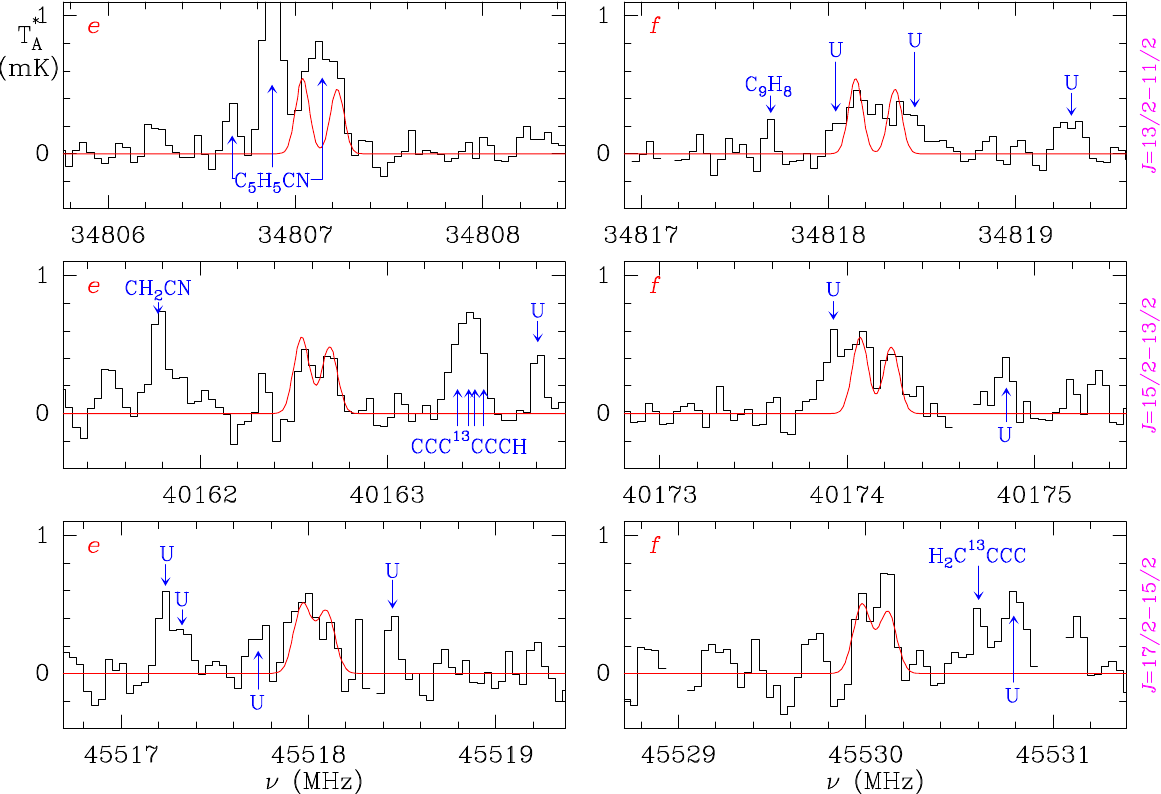}
\caption{Observed transitions of HC$_3$S. Quantum numbers are indicated at the
right side of each row.
The abscissa corresponds to the rest frequency. The ordinate is the antenna temperature, corrected for
atmospheric and telescope losses, in  millikelvins.
Blank channels correspond to negative features produced when folding the frequency-switched data.
The $\Lambda$-doubling components $e$ and $f$ for each $J$ are shown
in the left and right panels, respectively. In addition, the two hyperfine components of each 
$e$ and $f$ rotational transition 
are resolved in our data and appear clearly detected. They correspond to 
$F_u$=$J_u$+1/2 and $J_u$-1/2.
The red line shows the modelled spectra for HC$_3$S with the physical parameters given in Sect. \ref{sec:HCCCS}.}
\label{fig_HCCCS}
\end{figure}

\subsection{Discovery of HC$_3$S}\label{sec:HCCCS}

HCCS and HC$_4$S have been detected in TMC-1 \citep{Cernicharo2021b,Fuentetaja2022}; hence, we could expect HC$_3$S to
also be present in the cloud.
Rotational spectroscopy for HC$_3$S is available from the studies of
\citet{McCarthy1994} and \citet{Hirahara1994}. The dipole moment
was estimated to be 1.76\,D through ab initio calculations \citep{McCarthy1994}. Frequency predictions for this species are 
available in the CDMS catalogue; hence, we implemented them into
the MADEX code to obtain frequency predictions for the rotational
transitions inside the frequency coverage of QUIJOTE. Three of
these transitions -- with $J_{\rm up}$=13/2, 15/2, and 17/2 -- are in
our data.

The lines of
the $^2\Pi_{1/2}$ ladder of HC$_3$S exhibit $\Lambda$-doubling splitting that produces two components, $e$ and $f$, for each
rotational transition. In addition, each of these $e$ and $f$ components is further split into two hyperfine components due to the spin 1/2 
of the unpaired electron. Hence, the three rotational transitions of HC$_3$S that can be observed in our data will produce a 
total of 12 well-resolved features. Figure \ref{fig_HCCCS} shows the
observed lines. Although the $e$ component of the $J$=13/2-11/2 transition is partially blended with C$_5$H$_5$CN, the two hyperfine
components can be seen and adjusted. All the other lines are well
detected. However, due to the partial blending between the $F_u=J_u+1/2$ and
$F_u=J_u-1/2$ hyperfine components, a fixed linewidth of 0.7 km\,s$^{-1}$ was adopted for all the lines.
We conclude that we have detected enough individual lines of HC$_3$S in TMC-1 to confidently claim its first 
detection in space.
The derived line parameters are given in Table \ref{line_parameters_HCCCS}.

We adopted the same physical parameters for the cloud as for NC$_3$S and derive a rotational temperature  of 6.0$\pm$0.5\,K and
a column density of (1.5$\pm$0.2)$\times$10$^{11}$ cm$^{-2}$, which is roughly a factor of two below the upper
limit derived by \citet{Cernicharo2021b}. The HCS and HCCS column densities
have been derived to be (5.5$\pm$0.5)$\times$10$^{12}$ and (6.8$\pm$0.6)$\times$10$^{11}$ cm$^{-2}$, 
respectively \citep{Cernicharo2021b}.
HC$_4$S was recently discovered towards TMC-1 by \citet{Fuentetaja2022} with an estimated column density 
of (9.5$\pm$0.8)$\times$10$^{10}$ cm$^{-2}$. Consequently, we derive the following abundance ratios:
HCS/HCCS\,=\,8.1$\pm$1.6, HCCS/HC$_3$S\,=\,4.5$\pm$1.0, and HC$_3$S/HC$_4$S\,=\,1.6$\pm$0.3. It seems, hence, that longer
members of the HC$_n$S family could be present in significant abundances in TMC-1. 

HC$_5$S was observed in the laboratory by \citet{Gordon2002}.
Frequency predictions for this species are included in the CDMS catalogue and have 
been used to implement the molecule in MADEX. 
Adopting an abundance ratio between HC$_4$S and HC$_5$S similar to that between HC$_3$S and HC$_4$S, we could
expect a column density for HC$_5$S of $\sim$6$\times$10$^{10}$ cm$^{-2}$. For a rotational temperature of 6\,K,
the expected line intensities for the $J$=37/2-35/2 transition at 32.4 GHz are $\sim$0.2\,mK. 
These intensities are just the $3\sigma$ level of QUIJOTE, and, unfortunately, none of the lines of HC$_5$S have 
been detected. The same applies to longer members of the HC$_n$S family.

\begin{table}
\centering
\tiny
\caption{Estimated line parameters for the observed lines of HC$_3$S.}
\label{line_parameters_HCCCS}
\begin{tabular}{cccccc}
\hline
Transition~$^a$&$\nu_{cal}$~$^b$& $\int T_A^* dv$~$^c$ & v$_{LSR}$~$^d$  & T$_A$$^*$$^e$ & Notes\\
 ($J_u,p,F_u$) &(MHz)          & (mK km\,s$^{-1}$)    & (km\,s$^{-1}$)&  (mK)         & \\
\hline
13/2,e,7& 34807.038& 0.42$\pm$0.08         &5.72$\pm$0.10& 0.56$\pm$0.07& A  \\ 
13/2,e,6& 34807.223& 0.49$\pm$0.07         &5.86$\pm$0.07& 0.66$\pm$0.07& A  \\ 
13/2,f,7& 34818.146& 0.40$\pm$0.05         &5.73$\pm$0.08& 0.51$\pm$0.07& B  \\ 
13/2,f,6& 34818.356& 0.40$\pm$0.05         &5.57$\pm$0.10& 0.37$\pm$0.07& B  \\ 
15/2,e,8& 40162.543& 0.33$\pm$0.08         &5.67$\pm$0.08& 0.49$\pm$0.08&   \\ 
15/2,e,7& 40162.693& 0.34$\pm$0.09         &5.82$\pm$0.08& 0.45$\pm$0.08&   \\ 
15/2,f,8& 40174.070& 0.51$\pm$0.06         &5.77$\pm$0.05& 0.69$\pm$0.10& C  \\ 
15/2,f,7& 40174.236& 0.40$\pm$0.06         &5.79$\pm$0.06& 0.53$\pm$0.10&   \\ 
17/2,e,9& 45517.969& 0.27$\pm$0.06         &5.88$\pm$0.14& 0.36$\pm$0.13&   \\ 
17/2,e,8& 45518.095& 0.47$\pm$0.06         &5.89$\pm$0.09& 0.63$\pm$0.13&   \\ 
17/2,f,9& 45529.980& 0.33$\pm$0.10         &5.83$\pm$0.09& 0.60$\pm$0.13&   \\ 
17/2,f,8& 45530.118& 0.56$\pm$0.11         &5.90$\pm$0.07& 0.80$\pm$0.13&   \\ 
\hline
\end{tabular}
\tablefoot{
\tablefoottext{a}{Rotational upper quantum numbers. All transitions correspond to $\Delta$$J$=$\Delta$$F$=+1.}
\tablefoottext{b}{Predicted frequencies from a fit to the laboratory data  \citep{McCarthy1994,Hirahara1994}. The uncertainty for all frequencies is 2 kHz.}
\tablefoottext{c}{Integrated line intensity in mK\,km\,s$^{-1}$.}
\tablefoottext{d}{Derived velocity with respect to the standard local of rest (LSR) in km\,s$^{-1}$.}
\tablefoottext{e}{Antenna temperature in millikelvins.}
\tablefoottext{A}{Partially blended with a line from C$_5$H$_5$CN. The line is between the two
hyperfine components of this transition of HC$_3$S.}
\tablefoottext{B}{Partially blended with two unknown features. The derived line parameters are 
uncertain.}
\tablefoottext{C}{This line is partially contaminated by an unknown feature at a velocity of 6.9 km\,s$^{-1}$.
The unknown and HC$_3$S features can be fitted reasonably well.}
}
\end{table}
\normalsize

\section{Discussion}\label{sec:discussion}

\begin{figure}
\centering
\includegraphics[width=\columnwidth]{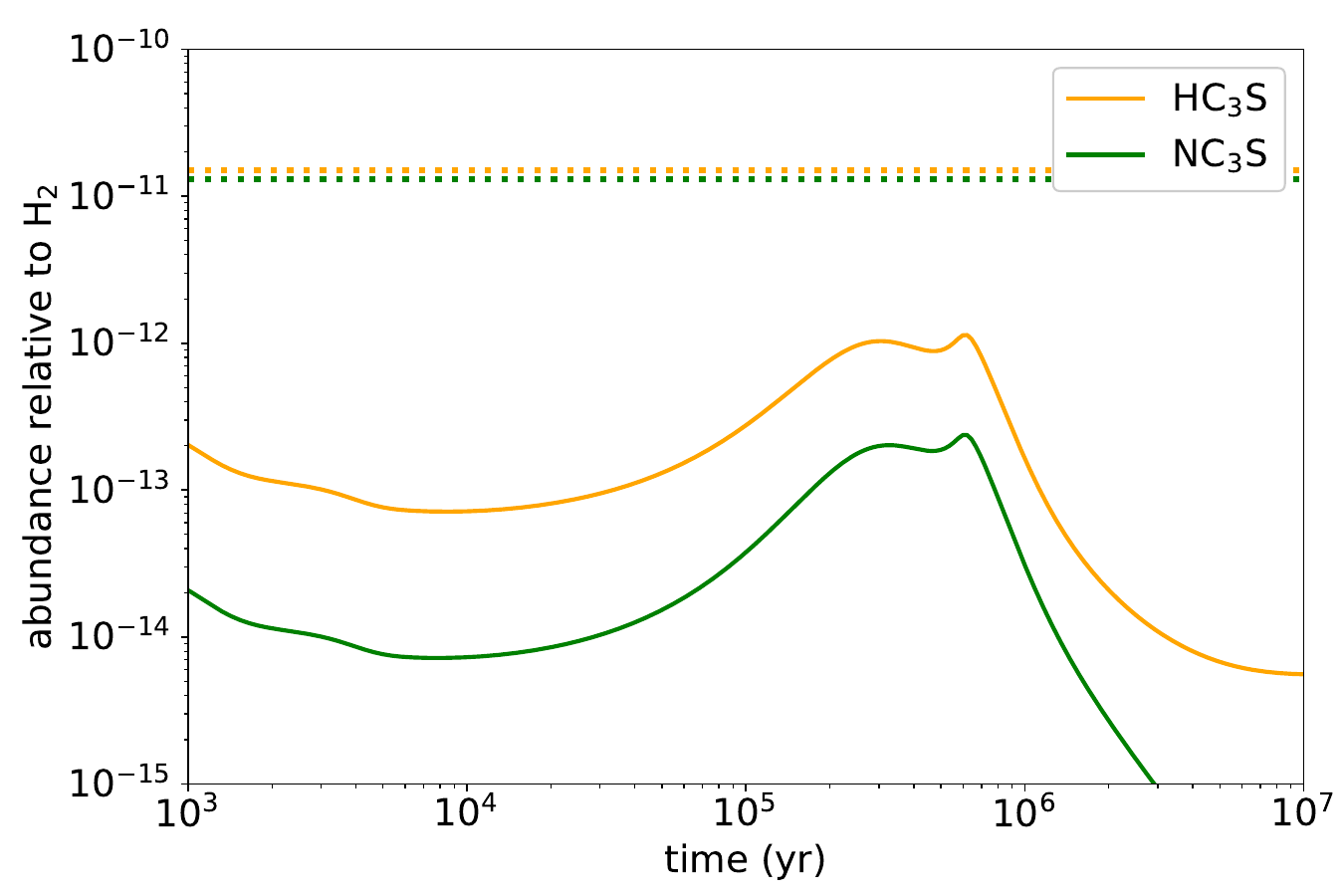}
\caption{Calculated fractional abundances of HC$_3$S and NC$_3$S as a function of time. Horizontal dotted lines correspond to 
the values observed in TMC-1 when adopting a column 
density of H$_2$ of 10$^{22}$ cm$^{-2}$ \citep{Cernicharo1987b}.}
\label{fig:abun}
\end{figure}

The chemistry of sulphur-bearing molecules in cold dark clouds
has recently been studied by \cite{Vidal2017}, \cite{Vastel2018}, and \cite{Laas2019} based on new chemical network developments. These studies reveal that the chemistry of sulphur in these cold clouds depends strongly on the degree of depletion of this element onto dust grains. Chemical networks are rather incomplete when dealing with
S-bearing species. For example, of the first species detected in TMC-1 with QUIJOTE (NCS, HCCS, H$_2$CCS, H$_2$CCCS, C$_4$S, and C$_5$S), only C$_4$S was included in the chemical networks
RATE12 \citep[UMIST;][]{McElroy2013} and kida.uva.2014 \citep[KIDA;][]{Wakelam2015}. 
\citet{Vidal2017} made a significant effort
to expand the number of reactions involving S-bearing species,
including some of the species that were later detected with QUIJOTE. The latest chemical network release of the UMIST 
database \citep{Millar2024} included many of the reactions first introduced by \cite{Vidal2017}.

To investigate the formation mechanism of HC$_3$S and NC$_3$S in TMC-1, we carried out chemical 
modelling calculations by adopting physical parameters of a standard cold dense cloud \citep{Agundez2013} and
using the UMIST 2022 chemical network \citep{Millar2024}. The number of reactions involving our newly detected species is very limited. HC$_3$S appears as reactant in 16 reactions and as a product in only 10 reactions, and NC$_3$S was not included. According to our calculations, the main formation reaction of HC$_3$S is S + C$_3$H$_2$, followed by the dissociative recombination of HC$_3$SH$^+$, and the reaction between C and H$_2$CCS. We included additional neutral--neutral reactions assuming that they occur through H atom elimination with a rate coefficient of 10$^{-10}$ cm$^3$ s$^{-1}$ (CH$_2$ + C$_2$S, CH + HCCS, SH + C$_3$H, and C$_2$H + HCS), but they are not as efficient as S + C$_3$H$_2$. In the model, HC$_3$S is mainly destroyed through reactions with neutral atoms (O, N, and C), which are assumed to occur fast according to \cite{Vidal2017}. The resulting peak abundance calculated for HC$_3$S lies one order of magnitude below the observed one (see Fig.\,\ref{fig:abun}). It is unclear whether the chemical model is missing reactions of formation or if it is overestimating the destruction via reactions with neutral atoms.

The chemistry of NC$_3$S is completely unexplored. To shed some light on the possible formation mechanism of this radical, we included several neutral--neutral reactions that could lead to NC$_3$S, adopting a rate coefficient of 10$^{-10}$ cm$^3$ s$^{-1}$. These reactions are N + HC$_3$S, C$_3$N + SO, C$_3$N + S$_2$, C$_3$N + SH, and NH + C$_3$S. Based on the main destruction processes of NCS (which is included in the UMIST 2022 network), we assumed that NC$_3$S is mainly destroyed through reactions with O, H, and C atoms, and via reactions with abundant cations such as HCO$^+$, H$_3$O$^+$, and H$_3^+$. The calculated abundance of NC$_3$S is shown in Fig.\,\ref{fig:abun}, which shows that the peak calculated abundance lies 50 times below the observed value. The main formation reaction of NC$_3$S is N + HC$_3$S. Since in the model HC$_3$S is the main precursor of NC$_3$S, the abundance of the latter closely follows that of the former (see Fig.\,\ref{fig:abun}).

Taking into account the reduced set of reactions involving HC$_3$S and NC$_3$S that are included in the chemical model and the large uncertainties in the reaction rate coefficients, the predictions of the chemical model regarding these two molecules should be viewed with caution, just as an order of magnitude estimate. A careful study of the main reactions that are identified here to most influence these two molecules, namely S + C$_3$H$_2$ and N + HC$_3$S, together with the reactions of HC$_3$S and NC$_3$S with neutral atoms would provide a more accurate view of the underlying chemical processes responsible for the presence of these two peculiar S-bearing molecules in the cold dark cloud TMC-1.

\begin{acknowledgements}

We thank Ministerio de Ciencia e Innovaci\'on of Spain (MICIU) for funding support through projects 
PID2019-106110GB-I00, and PID2019-106235GB-I00. We also thank ERC for funding through grant 
ERC-2013-Syg-610256-NANOCOSMOS. We thank the Consejo Superior de Investigaciones Cient\'ificas 
(CSIC; Spain) for funding through project PIE 202250I097.

\end{acknowledgements}

\normalsize

\end{document}